\input{epsf}
\documentstyle[prl,aps]{revtex}
\begin{document}
\draft
\title{Gravitational waves and pulsating stars: 
What can we learn from future observations?}

\author{Nils Andersson$^{1,2}$ and Kostas D. Kokkotas $^{3,4}$}

\address{$^1$ Department of Physics,  Washington University, St Louis MO 63130, 
USA}
\address{$^2$ Department of Physics and Astronomy,
University of Wales College of Cardiff,
Cardiff CF2 3YB, United Kingdom}  
\address{$^3$ Max-Planck-Society Research-Unit Theory of Gravitation, 
University of Jena, D-07743 Jena, Germany}
\address{$^4$ Department of Physics, Aristotle University of Thessaloniki,
Thessaloniki 54006, Greece}

\twocolumn[
\maketitle

\begin{abstract}
\widetext
We present new results for pulsating stars in general
relativity. First we show that the so-called gravitational-wave
modes of a neutron star can be excited when a gravitational wave 
impinges on the star. Numerical simulations suggest that the modes may be
astrophysically relevant, and we discuss whether they will be
observable with future gravitational-wave detectors. 
We also discuss how such observations 
could lead to estimates of both the radius and the mass of 
a neutron star, and thus put constraints on the nuclear equation of state.
\end{abstract}

]

\narrowtext

Pulsating stars have been studied within Einstein's
general theory of relativity for thirty years. 
Hence it is somewhat surprising that recent results have altered
our understanding of these problems considerably. 
We now know that there are oscillation modes that are directly
associated with the curvature of spacetime. But it is not yet clear
whether these modes are of any relevance to astrophysics, or if they 
are simply a curiosity of the mathematical theory. In order to
demonstrate their relevance one must show that the modes
are excited in a realistic
astrophysical situation, such as gravitational collapse to form 
a neutron star. Unfortunately, we do not yet have the ability to 
perform such calculations in a relativistic framework. 
While waiting for full 3D numerical relativity to provide a 
definitive statement regarding the relevance of
the gravitational-wave modes of a compact star, we must settle for 
suggestive calculations. The purpose of this letter is to describe one 
such calculation: We consider the scattering of gravitational-wave packets
off a uniform density star. In this case we 
find that the gravitational-wave modes 
are indeed excited.

Recent work on the pulsation properties of compact stars
has shown that general relativity plays an important role: A 
relativistic star
exhibits two distinct sets of pulsation modes. One is slowly damped and 
corresponds to the well-known fluid modes 
\cite{gau95,thorne67,lindblom83}, while the second is rapidly 
damped and has no analogue in Newtonian 
theory. 
The new modes have been termed w-modes because they are 
closely associated with gravitational waves 
\cite{kokkotas92}. 
We have recently managed to 
put the w-modes in context and provide a better 
understanding of the physics involved \cite{andersson95b,ica,kreview}. 

The equations describing perturbed nonrotating stars in general 
relativity split 
into two classes \cite{thorne67}. Polar perturbations correspond 
to zonal compressions of the star, whereas axial ones induce differential rotation
in the fluid. 
In general, the polar problem corresponds to two coupled wave equations: one 
that represents the fluid motion and one for the gravitational waves
\cite{Ipser}. The fluid pulsation modes (eg. the f-mode) belong to 
the polar class of perturbations.
Meanwhile, axial
perturbations are described by a single, homogeneous wave equation
\cite{thorne67}. 
The axial problem is thus considerably simpler. Nevertheless, it
attained very little attention until recently.  This is mainly
because an axial perturbation can only induce pulsations in the 
stellar fluid through a non-zero shear modulus \cite{Schumaker}, or through
coupling to the polar perturbations if the 
star is rotating \cite{cf91}. 
However, the axial problem is interesting also for the simplest 
stellar models: If the 
star can be made ultracompact ($R<3M$ in geometrical units $c=G=1$) the
peak of the exterior curvature potential barrier (that is familiar from black-hole
perturbation theory) will be unveiled. Then gravitational waves that impinge
on the star can be trapped. That such trapped modes exist has been 
demonstrated \cite{chandra91b,kokkotas94}, but they are unlikely to be of any 
great astrophysical
relevance: A sufficiently compact star will probably never form. 

More importantly, one can argue that the axial problem is  
relevant also for less compact stars (the canonical 
values $R=10$ km and $M=1.4M_\odot$ for  neutron stars lead to 
$M/R \approx 0.2$), 
especially as far as gravitational waves are concerned. For 
black holes, when the gravitational
waves are the only active agent, the axial problem leads to a
pulsation spectrum that is identical to the polar one.
Knowing this, the results of a detailed mode-survey for uniform 
density stars do not come as a complete surprise: 
The axial and the polar w-mode spectra are 
remarkably similar \cite{andersson95b}. 
This is an important result because of its 
implications for physical interpretations of the w-modes. Since the 
axial modes do not couple to the stellar fluid one
cannot invoke the 
fluid in an explanation of them. The w-modes must 
arise because gravitational waves can be temporarily 
trapped in the ``bowl of spacetime curvature'' provided by the mass
of the star \cite{andersson95b}.

{\em Are the w-modes excited?}. --- 
Our understanding of the role of a dynamic spacetime for 
pulsating stars has been improved, but the most important question 
remains. As yet it has not been 
established that the w-modes can contribute to observable 
gravitational waves, and thus play a role in astrophysics. 
The  modes should be excited 
when a neutron star is formed 
through gravitational collapse, or when two 
neutron stars merge at the final stage of a binary systems evolution, but
how much energy goes into the pulsation?
At the present time we cannot answer such questions: To perform the
required calculations in a relativistic framework is 
simply beyond our means.

As a first step towards understanding the issue
we have studied scattering of 
axial gravitational wave-packets by a compact star. 
Since axial perturbations are governed by a single wave 
equation with an 
effective potential \cite{thorne67,chandra91b}, this problem is in many 
respects identical 
to that for black holes \cite{vishu}. 

\begin{figure}
\epsfxsize=250pt \epsfbox{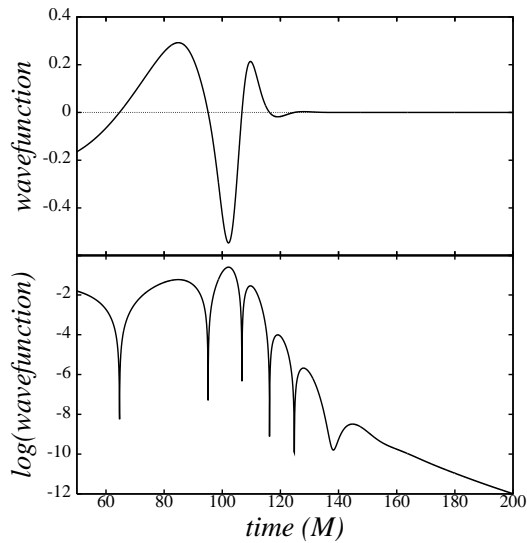}
\caption{The response of a uniform density star ($M/R = 0.2$) to a Gaussian pulse 
of axial gravitational waves. The top panel shows the axial 
perturbation as seen by a distant observer, while the lower panel shows 
the same 
function on a logarithmic scale.}
\end{figure}

The result of a typical simulation is shown in Figure 1.
The exponential ringdown at late times  (from $t\approx 100M$) 
corresponds to the 
first axial w-mode. A power spectrum 
for the data in Figure 1 shows that the first
three axial modes are excited. A significant part of the initial energy 
clearly goes into quasinormal-mode ringing. 
To study the relation between the fluid modes and the w-modes we must turn
to the polar problem. Preliminary studies \cite{aaks} indicate that i)  a 
polar perturbation typically excites both the fluid modes and the w-modes
and ii) a considerable amount of energy is released through the w-modes.

Our simulations suggest that the 
w-modes may be relevant in many dynamical processes involving neutron stars. 
We believe that the modes ought to be excited 
when a stellar core collapses to form a neutron star. Most of the initial 
deformation of spacetime could then be radiated away in terms of w-modes.
For the mode-excitation to be considerable the collapse must, of course, be 
asymmetric. To estimate the asymmetries in a realistic supernova 
is difficult, but the evidence is that the average velocity of
radio pulsars is large. Asymmetries in the core collapse would explain such
high velocities \cite{Burrows}.  Moreover, since it is the asymmetries of the spacetime that are
relevant for the excitation of the w-modes the situation for these modes is  not too different from that for black holes. It is well-known that both axial
and polar quasinormal modes are generic features of collapses that
form a black hole \cite{seidel}, and that
the radiation following the collision of two black holes is
dominated by quasinormal-mode ringing \cite{2bh}. Hence, it does not 
seem unreasonable to assume that 
the  w-modes will be excited when a neutron star is 
created or when two such stars collide.

One important conclusion that follows from our simple simulations is that 
{\em studies of
dynamical processes involving neutron stars must be performed using
general relativity}. This may seem obvious, but at the present time a
considerable effort is invested in calculations using Newtonian gravity.
The gravitational radiation is typically extracted by means of the 
quadrupole formula.
The w-modes will never appear in such calculations and thus this feature
of the full problem will be unaccounted for.  
 
It is clear that the stellar pulsation 
modes can play a role in many astrophysical 
scenarios, but will we be able to observe them with future gravitational-wave
detectors? In the high-frequency regime where both the fluid f-mode and the w-modes reside
(1-2 kHz and 8-12 kHz, respectively) one would not expect too much from the
new generation of laser-interferometric gravitational-wave detectors. 
These will probably not be sensitive enough at 
such high frequencies (although their performance in this regime may be much improved 
through dual recycling). But neutron stars pulsating in the f-mode 
are ideal sources for the 
resonant bar detectors that are currently operating. 
The frequency of the f-mode 
also makes it well suited for detection by 
the recently proposed 
spherical solid-mass detectors, {\em e.g.}, the Transverse Icosahedral
Gravitational Wave Antenna (TIGA) \cite{tiga}.
Meanwhile, the w-modes provide interesting sources for the 
 proposed array of smaller bar detectors, 
which should be sensitive in the few kHz regime \cite{papa}.

To obtain rough estimates for the typical gravitational-wave amplitudes from a pulsating star we use the standard relation for the 
gravitational-wave flux \cite{bfs}
\begin{equation}
F = {c^3 \over 16\pi G} \vert \dot{h} \vert = 
{1\over 4\pi r^2 } {dE \over dt} ,
\end{equation}
which is valid far away from the star. Combining this with i)
${dE/dt} = E/2\tau$ where $\tau$ is the e-folding time of the pulsation
and $E$ is the available energy ii) the assumption 
that the signal is monochromatic (with frequency $f$) and iii) the knowledge
that the effective amplitude achievable after matched filtering scales
as the square-root of the number of observed cycles:
$h_{\rm eff} = h\sqrt{n} = h\sqrt{f\tau}$, we get the estimates
\begin{equation}
h_{\rm eff} \sim 3\times 10^{-21} \left( {E\over 10^{-6} M_\odot c^2}
 \right)^{1/2} \left(  { 2 {\rm kHz} \over f } \right)^{1/2}
\left( {50{\rm kpc} \over r} \right) \ ,
\end{equation}
for the f-mode, and
\begin{equation}
h_{\rm eff}\sim 1\times 10^{-21} \left( { E \over 10^{-6} M_\odot c^2}
 \right)^{1/2} \left( { 10 {\rm kHz} \over f } \right)^{1/2}
\left( { 50{\rm kpc} \over r } \right) \ ,
\end{equation}
for the fundamental w-mode.
Here we have used typical parameters for the pulsation modes,
and the distance scale used is that to SN1987A.
In this volume of space one would not expect to see more than 
one event per ten years or so.  
However, the assumption that the  energy release in 
gravitational waves in a supernova is of the order of 
 $10^{-6} M_\odot c^2$ is very conservative \cite{bfs}. 
If 
a substantial fraction of the binding energy of a neutron star 
were released through the pulsation modes
we could hope to see 
such events out to the Virgo cluster (and perhaps as many as a few per year). 
  
{\em What can we learn from observations?}. ---
If they were detected, the stellar pulsation modes 
carry key information about the nature 
of neutron stars. Suppose that we detect a gravitational-wave
signal from a compact star, what can we hope to learn from it? 
Again, a detailed answer requires much further study. But it is 
useful to speculate on the possibilities.

Assume that we  detect a signal and manage
to extract both the fluid f-mode and the slowest damped polar (or axial) 
w-mode 
from it. Then spectral studies suggest that:
i) the oscillation frequency of the f-mode scales with the average density 
of the star through $\sqrt{M/R^3}$, and ii) the damping rate of the w-mode depends linearly 
 on the compactness ratio  $M/R$ of the star.
These properties are illustrated in Figures 2a,b.
In principle, one should  be able to infer the average density of the 
star from an observation of the f-mode. Similarly, an observation of a 
w-mode leads to an estimate of the stars compactness. Although 
important, the information
available in either case would not provide detailed information about the 
stellar parameters. But the situation is markedly different if both 
modes are observed. Then \underline{both} 
the mass and the radius of the star can be extracted from the observed data.
This 
information puts strong constraints on the
nuclear equation of state \cite{lblom}.  

The idea seems simple enough, but will it be useful in practice?
Let us give an example for stars with polytropic equations of state:
We have constructed a set of independent polytropic stellar models ($p=K\rho^{1+1/n}$)
with varying polytropic index ($n=0.5;0.8;1;1.2;1.5$). We have  
determined the f-mode and the slowest damped
polar w-mode for each of these models (the results would be similar
if we use the fundamental axial w-mode). The relevant data is graphed
in Figures 2a,b. The theoretical spectra are well fitted by the following
two relations: The oscillation frequency of the f-mode 
varies with the average density of the star as
\begin{equation}
\omega_f ({\rm kHz}) \approx 0.17 + 
2.30 \sqrt{ \left( {10{\rm km} \over R} \right)^3  \left({M \over 1.4 M_\odot}\right) } \ ,
\label{detec1}
\end{equation}
while the damping rate of the w-mode behaves as ($\tau_\omega$ is the e-folding time of the pulsation)
\begin{equation}
{1 \over \tau_w} ({\rm MHz}) \approx 0.107 - 
0.069 \left( {10{\rm km} \over R} \right)  \left({M \over 1.4 M_\odot}\right) \ .
\label{detec2}
\end{equation}
These approximations are shown in Figures 2a,b.
The crucial question  is: How robust are these fits for manifestly
independent stellar models? The two relations (\ref{detec1}) and
(\ref{detec2}) must provide reasonably accurate estimates for 
both $M$ and $R$ for all stars in our dataset in order for the idea to 
be useful. The scheme passes this first simple
test with flying colours: By inverting (\ref{detec1}) and 
(\ref{detec2}), the mass and the radius of each star can be 
determined at least as accurately as can the average density and the 
stellar compactness. The error in $M$ and $R$ is, in fact, typically smaller 
than 10\% .

\begin{figure}
\epsfxsize=250pt \epsfbox{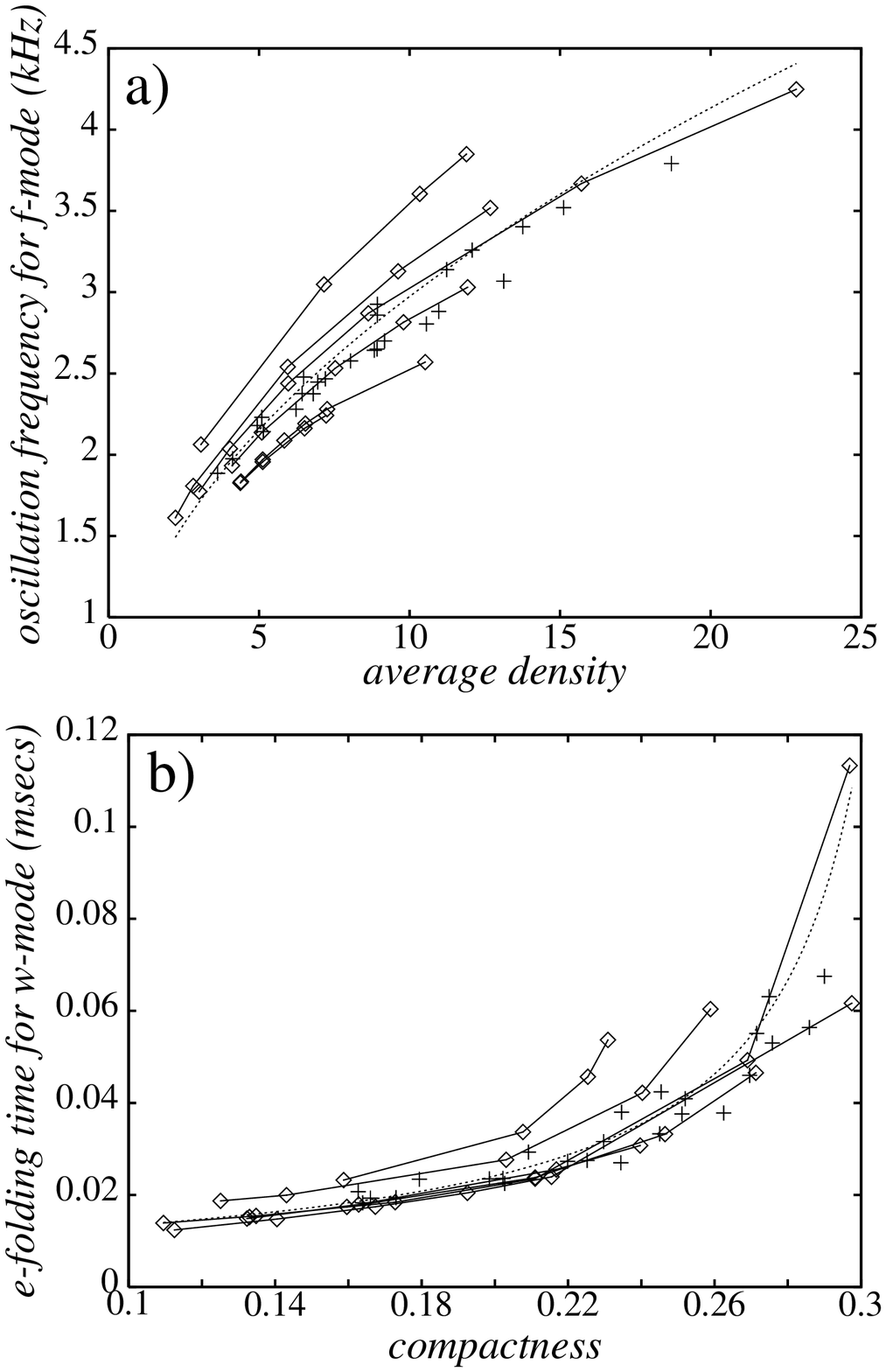}
\caption{ (a) The pulsation frequency of the f-mode as a function of the average density
(in units $10^{14}{\rm g/cm}^3$) of the star , and (b) the e-folding time 
for the slowest damped polar w-mode
as a function of the stellar compactness (dimensionless). The 
behaviour is not
strongly dependent on the polytropic model used, and therefore ideal
for parameter extraction. The polytrope data that is used in the 
discussion in the text is shown as diamonds connected with solid 
lines. The corresponding least-square fits are shown as dashed curves. 
Also indicated (by crosses) are preliminary results for a few 
realistic equations of state. }
\end{figure}

This is clear  evidence that our idea can be useful in a real 
detection situation, and that it deserves to be investigated in
more detail. One must: 
i) incorporate the estimated
effects of statistical errors and measurement ones. It will, for example,
be much more difficult to infer the w-mode damping rate from a data set than
to find the f-mode pulsation frequency (remember that 
the observability of a periodic
signal buried in noise scales roughly as the square-root of the number
of observed cycles).
ii) obtain fits similar to (\ref{detec1}) and (\ref{detec2})
for more realistic equations of state. If such relations prove 
as robust as the ones we have obtained for polytropes then the
suggested scheme looks truly promising.

We are presently investigating these questions and will
return to them in the future. As yet, we have done 
preliminary calculations for seven of the equations of state that 
were used by Lindblom and Detweiler \cite{lindblom83} 
(their models A, B, C, E, F, G and I). These results, which have 
been included in Figures 2a,b,  are in impressive agreement with the 
polytrope data used in our example. As a further demonstration of the
robustness of the suggested scheme we have considered the f-mode 
data that Cutler,
Lindblom and Splinter obtained for several realistic equations of state 
\cite{cutler90}. One can use the stellar 
parameters $M$ and $R$ from Table 2 in \cite{cutler90} together
with our relation (\ref{detec1}). This leads to approximate f-mode
frequencies that should 
be compared to the data in Table 3 of \cite{cutler90}.
Despite being derived for 
polytropes, our equation (\ref{detec1}) approximates almost all the 
$\ell=2$ frequencies in Table 2 of \cite{cutler90} to well within 10 \%.

{\em Concluding remarks}. --- In this letter we have presented 
suggestive results for problems that can have direct impact on the future
detection of gravitational waves from neutron stars. 

First we have shown that the gravitational-wave modes of a neutron
star will be excited in a dynamical situation. The modes should be 
excited when a neutron star is
formed in a gravitational collapse, 
a process when a considerable amount of energy is released. 
But at present it is difficult to estimate the amount of  energy 
that goes into the pulsation modes. 
What is absolutely clear (and this
is a very important conclusion) is that we are asking 
questions that can never be answered within Newtonian gravity. 
{\em The assertions presented here must be tested
by more detailed, fully general relativistic, simulations.}
This is a challenge for numerical relativity. 
A relativistic description of gravitational collapse to form a
neutron star, or the merger of two stars,  should tell us whether 
the w-modes are of observational relevance or not.
Furthermore,  the pulsation properties of a slightly perturbed
star can be used as a powerful test of the reliability of a numerical code.  

The second part of our work shows that the stellar pulsation modes 
can, if observed, 
provide us with accurate information about the stellar parameters.
We have illustrated how a simple scheme can lead to 
estimates of both the mass and the radius of the star. This is 
a very useful suggestion, since it can put
strong constraints on the nuclear equation of state. 

{\em Given the obvious importance for high-frequency
gravitational-wave detectors,  
further studies of the issues raised here are 
urgently required.}

We thank Bernard Schutz for many useful discussions and suggestions. 
We also thank Gabrielle Allen for help with the numerical 
work. This work was supported by an exchange program from the British Council 
and the Greek GSRT, and NA is supported by NFS (Grant no PHY 92-2290) and NASA (grant no NAGW 3874). KDK thanks MPG for generous hospitality.

\end{document}